\documentclass[10pt, conference]{IEEEtran}
\usepackage[utf8]{inputenc}
\usepackage[detect-all]{siunitx}
\usepackage{array,tabularx,calc}
\usepackage{mathtools}
\usepackage{amsfonts}
\usepackage{graphicx, balance}
\usepackage{amssymb}
\usepackage{pbox}
\usepackage{url}
\usepackage{algorithmic}
\usepackage[linesnumbered,ruled,vlined]{algorithm2e} 
\usepackage{tikz}
\usepackage{textcomp}
\usepackage{amsmath, amssymb}
\usepackage{comment}
\newcommand{\KwInput}[1]{\textbf{Input:} #1}
\newcommand{\KwOutput}[1]{\textbf{Output:} #1}

\DeclareSIUnit{\belmilliwatt}{Bm}
\DeclareSIUnit{\dBm}{\deci\belmilliwatt}
\DeclareSIUnit{\belisotropic}{Bi}
\DeclareSIUnit{\dBm}{\deci\belisotropic}
\DeclareSIUnit{\bit}{bit}

\ifCLASSOPTIONcompsoc
    \usepackage[caption=false,font=normalsize,labelfont=sf,textfont=sf]{subfig}
\else
    \usepackage[caption=false,font=footnotesize]{subfig}
\fi

\usepackage[capitalise]{cleveref}
\crefformat{equation}{(#2#1#3)}
\usepackage[noadjust]{cite}

\usetikzlibrary{arrows}
\usetikzlibrary{shapes}

\usepackage[a4paper,bindingoffset=0.2in,left=0.60in,right=0.60in,top= 0.8in,bottom=0.84in,footskip=.25in]{geometry}

\begin{document}

\title{Joint Optimization of Multi-UAV Deployment and 3D Positioning in Traffic-Aware Aerial Networks}

\author{\IEEEauthorblockN{Kamran Shafafi, Alaa Awad Abdellatif, Manuel Ricardo, Rui Campos }
    \IEEEauthorblockA{ 
     INESC TEC and Faculdade de Engenharia, Universidade do Porto, Portugal  \\
        \{kamran.shafafi, alaa.abdellatif, manuel.ricardo, rui.l.campos\}@inesctec.pt}} 
\maketitle

\begin{abstract}

Unmanned Aerial Vehicles (UAVs) have emerged as a key enabler for next-generation wireless networks due to their on-demand deployment, high mobility, and ability to provide Line-of-Sight (LoS) connectivity. These features make UAVs particularly well-suited for dynamic and mission-critical applications such as intelligent transportation systems and emergency communications. However, effectively positioning multiple UAVs in real-time to meet non-uniform, time-varying traffic demands remains a significant challenge, especially when aiming to optimize network throughput and resource utilization. In this paper, we propose an Efficient Multi-UAV Traffic-Aware Deployment (EMTAD) Algorithm, a scalable and adaptive framework that dynamically adjusts UAV placements based on real-time user locations and spatial traffic distribution. In contrast to existing methods, EMTAD jointly optimizes UAV positioning and minimizes the number of deployed UAVs, ensuring efficient UE-UAV association while satisfying the traffic demand of users.  
Simulation results demonstrate that EMTAD significantly improves network performance while reducing deployment overhead by minimizing the number of UAVs required in dynamic and traffic-aware environments.  
\end{abstract}

\begin{IEEEkeywords}
	Unmanned Aerial Vehicles, 
	UAV Placement, 
	Positioning Algorithm, 
	LoS Communications. 
\end{IEEEkeywords}

\section{Introduction}

The explosive growth of wireless data traffic, driven by bandwidth-intensive applications and the proliferation of connected devices, has created an urgent demand for flexible, high-capacity, and resilient communication infrastructures. As wireless networks progress toward 6G and beyond, there is a growing need for communication and sensing systems that are intelligent, adaptable, and energy-efficient. Unmanned Aerial Vehicles (UAVs) have emerged as a key enabler for augmenting terrestrial networks, offering rapid deployability, high mobility, and reliable Line-of-Sight (LoS) communication—particularly in complex or hard-to-reach environments. 
In particular, multi-UAV networks have shown great potential in extending coverage, offloading traffic, and enhancing throughput in dynamic or challenging environments \cite{10.1007/978-3-031-57523-5_19, 10980618}. These capabilities make UAVs highly valuable in diverse applications such as smart cities, disaster response, industrial automation, and surveillance.  
However, one of the key challenges in deploying multi-UAV networks lies in the optimal positioning of UAVs to adapt to real-time traffic variations and maximize network throughput. Static or naive positioning strategies can lead to inefficient resource utilization, coverage gaps, or overloaded nodes, especially in heterogeneous and non-uniform traffic scenarios. 

Recent studies have explored various strategies for multi-UAV positioning to enhance wireless network performance. For instance, \cite{shakhatreh2023efficient} proposed a two-stage algorithm combining K-means clustering and swarm optimization to optimize different network configurations, namely, the network-centric and user-centric approaches. Similarly, \cite{zhao2023clustering} introduced a user clustering and bandwidth optimization method to improve energy efficiency in UAV-assisted wireless communications. 
In dynamic environments, reinforcement learning has been employed to optimize UAV trajectories. For example, the authors in \cite{9963915} proposed a reinforcement learning-based method that jointly optimizes the trajectory and transmission power of UAVs, in order to maximize detection performance while meeting constraints on both the accuracy of estimation of radar parameters and the quality of communication. In their model, each UAV is equipped with a dual-function device capable of simultaneous sensing and communication. 
In \cite{9712375}, the authors proposed a distributed multi-agent reinforcement learning framework for UAV networks, where each UAV independently optimizes its position, transmit power, and sub-channel allocation using partial observations. 
However, these studies primarily focus on a fixed number of UAVs or static user distributions. None of the aforementioned works consider a comprehensive approach that minimizes the number of required UAVs while dynamically optimizing their positions based on real-time traffic distribution and user locations to maximize overall network throughput. 

\begin{figure}
	\centering
	\includegraphics[width=\linewidth, height=4cm]{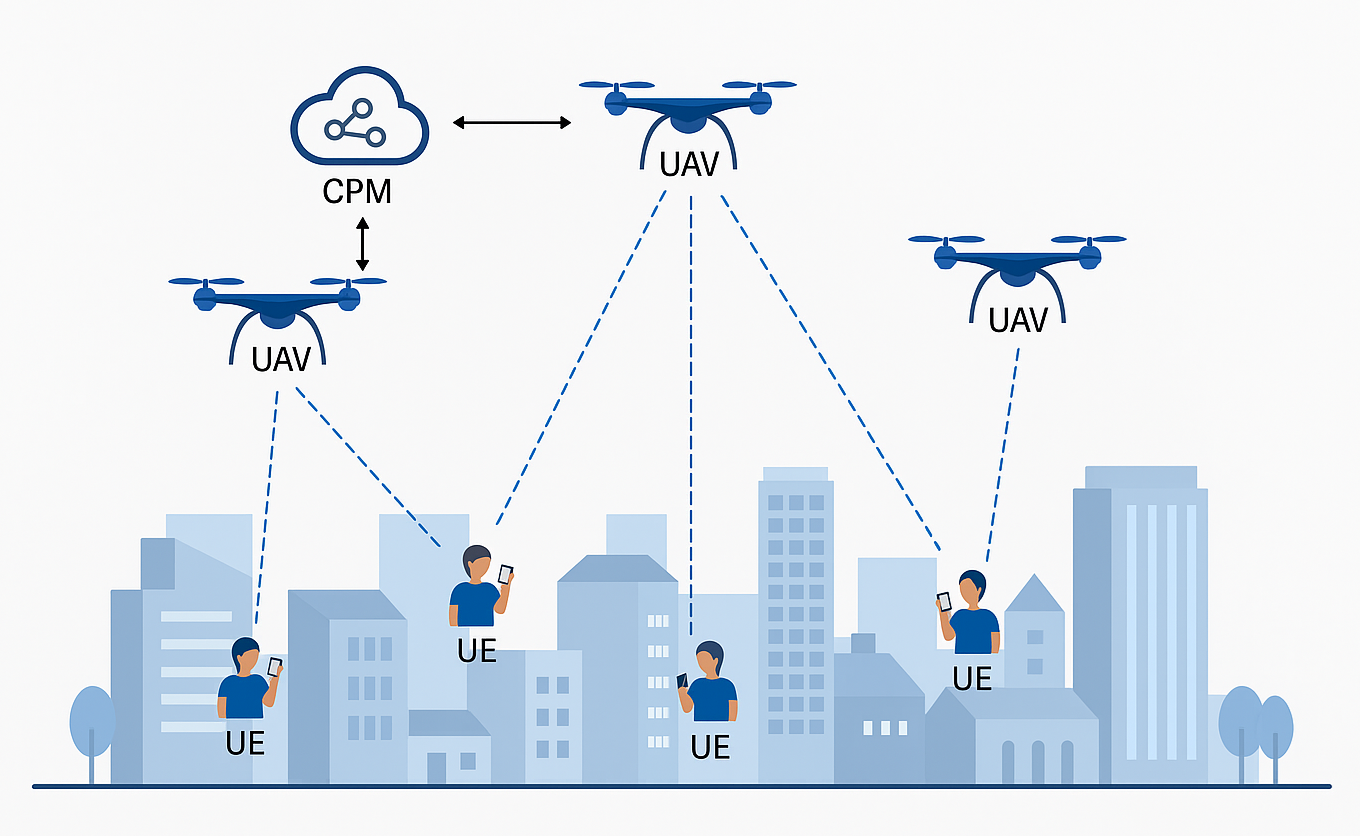}
	\caption{Multi-UAV system model under study.}
	\label{Fig:system}
\end{figure}

In this paper, we propose the Efficient Multi-UAV Traffic-Aware Deployment (EMTAD) algorithm—a novel framework aimed at minimizing the number of deployed UAVs while optimizing their 3D placement based on real-time traffic distribution and user association. Unlike prior works that assume static user patterns or a fixed UAV count, EMTAD dynamically adapts UAV deployment to spatial traffic variations, ensuring each user’s traffic demand is satisfied. This approach enhances LoS coverage, balances network load, and improves spectral efficiency. Thus, our main contributions are: (i) a traffic-aware, real-time UAV positioning strategy; (ii) an integrated optimization framework that jointly minimizes UAV count, determines UE-UAV associations, and accounts for practical deployment constraints; and (iii) a scalable algorithm designed for dynamic, heterogeneous network environments. This work builds upon our prior research in~\cite{10792915, 10539557}, which focused on single-UAV positioning but did not address the challenges of multi-UAV optimization or dynamic user association.

The following sections present our system model and problem formulation (Section \ref{sec:System}), the proposed Efficient Multi-UAV Traffic-Aware Deployment (EMTAD) algorithm (Section \ref{Traffic- and Obstacle-aware Positioning Algorithm}), the performance evaluation results (Section \ref{sec:performance_evaluation}), and the conclusions (Section \ref{sec:conclusions}).

\section{System Model and Problem Formulation} \label{sec:System}

In this section, we first introduce the considered system model, followed by the formulation of the joint optimization problem that governs UAV deployment and 3D positioning.

\subsection{System Model}

This paper considers a multi-UAV-assisted aerial wireless communication system deployed over a target area categorized as urban, suburban, residential, or rural (see Figure \ref{Fig:system}\footnote{This image was generated using ChatGPT (OpenAI, 2025) based on the author's conceptual input for illustrative purposes.}). The system comprises the following key components:\\
\textbf{i) User Equipment (UEs):} A total of \( N \) UEs are randomly distributed across the venue. Each UE is served by at least one of the \( K \) UAVs acting as aerial access points. \\    
\textbf{ii) UAVs:} Initially, the number of UAVs \( K_\text{max} \) is set equal to \( N \). The system then optimizes and minimizes \( K_\text{max} \) while ensuring a full UE-UAV association, so that it satisfies their traffic demands. UAVs are positioned to provide high-capacity directional links while preserving LoS with the served UEs. \\    
\textbf{iii) Central Processing Module (CPM):} A centralized cloud-based controller oversees the UAV network. Its main functions include:
(i) solving the formulated optimization problem that aims to minimize the number of deployed UAVs while fulfilling users' demand constraints; and (ii) obtaining the optimal 3D positions for each UAV and sending repositioning commands accordingly.\\

\subsection{Problem Formulation} \label{sec:problem_formulation}

This section presents a joint optimization problem that aims to minimize the number of deployed UAVs, determine their optimal 3D positions, and associate ground UEs with UAVs, all while satisfying traffic demands and resource constraints. The objective is to ensure reliable and efficient wireless connectivity for ground users in urban environments. 
The system operates in discrete time slots \( \tau \), during which UE locations are assumed static, while UAV positions are dynamically optimized. The aerial network is modeled as a directed graph \( \mathcal{G}(\tau) = \langle \mathcal{U}, \ell(\tau) \rangle \), where \( \mathcal{U} \) includes \( N \) UEs (\( \mathrm{UE}_i, \ i = 1, \ldots, N \)) and \( K_\text{max} \) UAVs (\( \mathrm{UAV}_k, \ k = 1, \ldots, K_\text{max} \)), and \( \ell(\tau) \) denotes the set of wireless links at time \( \tau \).  
Each \( \mathrm{UE}_i \) is located at fixed coordinates \( (x_i, y_i, z_i) \), typically with \( z_i \approx 0 \), and is associated with a time-varying traffic demand \( T_i(\tau) \) in bits per second. This demand captures the data rate required for uplink transmission to the serving UAV, and is application-dependent (e.g., video streaming, real-time sensing).

Inspired by~\cite{zhang2023trajectory}, the LoS probability between \( \mathrm{UE}_i \) and \( \mathrm{UAV}_k \) at position \( P_k = (x_k, y_k, z_k) \) is formulated as: 
\begin{equation}\label{eq:los_prob}
\varepsilon_{ik}^{\text{LoS}}( \tau) = \frac{1}{1 + c_1 \exp \left( -c_2 \left( \frac{180}{\pi} \arcsin \left( \frac{z_k - z_i}{d_{ik}( \tau)} \right) - c_1 \right) \right)}. 
\end{equation}
Hence, the probability of Non-Line-of-Sight (NLoS) is given by: 
\[
\varepsilon_{ik}^{\text{NLoS}}( \tau) = 1 - \varepsilon_{ik}^{\text{LoS}}( \tau)
\]
where \( c_1 \) and \( c_2 \) are environmental constants derived from the Al-Hourani model for urban settings~\cite{AlHourani2014}, and
\begin{equation}\label{diks}
 d_{ik}( \tau) = \sqrt{(x_k - x_i)^2 + (y_k - y_i)^2 + (z_k - z_i)^2}   
\end{equation}

is the Euclidean distance between \( \mathrm{UE}_i \) and \( \mathrm{UAV}_k \). These constants depend on the environment (e.g., urban, suburban, rural), reflecting the impact of building density, height, and terrain on LoS likelihood. 
The average channel gain is defined as:
\begin{equation}\label{eq:channel_gain}
\Omega_{ik}( \tau) = K_0^{-1} d_{ik}( \tau)^{-2} \left[ \varepsilon_{ik}^{\text{LoS}}( \tau) \mu^{\text{LoS}} + \varepsilon_{ik}^{\text{NLoS}}( \tau) \mu^{\text{NLoS}} \right]^{-1}
\end{equation}
where \( K_0 = \left( \frac{4\pi f}{c} \right)^2 \), and \( \mu^{\text{LoS}}, \mu^{\text{NLoS}} \) are attenuation factors~\cite{zhang2023trajectory}. 
Hence, the data rate from \( \mathrm{UE}_i \) to \( \mathrm{UAV}_k \) is given by:
\begin{equation}
R_{ik}( \tau) = B_i( \tau) \log_2 \left( 1 + \frac{P_T G_t G_r\Omega_{ik}( \tau)}{N_0 B_i( \tau)} \right)
\label{eq:datarate}
\end{equation}
where \( B_i(\tau) \) is the allocated bandwidth for $UE_i$.  

Our goal is to formulate a unified optimization problem that simultaneously minimizes the number of UAVs, determines their optimal 3D positions, and establishes efficient UE-UAV associations, while ensuring that each user's data demand is met. This approach aims to balance the trade-off between resource efficiency and overall network performance. Furthermore, we define a binary UAV activation variable \( a_k \), indicating whether $UAV_k$ is active (deployed).   
Accordingly, the optimization problem is formulated as follows:  

{ \small
\begin{subequations}\label{opt_integrated}
\begin{flalign}
\mbox{\bf P:} &  \min_{a_k, z_{ik}, P_k} \quad \sum_{k=1}^{K_\text{max}} a_k  & \label{eq:obj}  \\[1ex]
&\text{subject to:} & \nonumber \\[1ex]
&  R_{ik}( \tau) \geq  z_{ik} T_i(\tau), \ \  \forall i \in \{ 1, \ldots, N\}, \ \forall k \in \{ 1, \ldots, K_\text{max} \}, & \label{eq:demand} \\[1ex] 
& \sum_{i=1}^N z_{ik} \cdot  B_i(\tau) \leq B^{\text{MAX}}_k,  \quad \forall k \in \{1, \ldots, K_\text{max} \}, & \label{eq:capacity} \\[1ex]
& \sum_{k=1}^{K_\text{max}} z_{ik} = 1, \quad \ \  \forall i \in \{ 1, \ldots, N\}, & \label{eq:coverage} \\[1ex] 
& z_{ik} \leq a_k, \quad \quad \quad    \forall i \in \{ 1, \ldots, N\}, \ \forall k \in \{ 1, \ldots, K_\text{max} \}, &  \label{eq:Link} \\[1ex]  
& z_{ik} \in \{0, 1\}, \quad \ \  \forall i \in \{ 1, \ldots, N\}, \ \forall k \in \{ 1, \ldots, K_\text{max} \}, &  \label{eq:binary} \\[1ex] 
& a_{k} \in \{0, 1\}, \quad \quad   \forall k \in \{ 1, \ldots, K_\text{max}\}, &  \label{eq:activation} 
\end{flalign}
\end{subequations}
} 

\noindent where \( K_\text{max}=N \) denotes the maximum number of UAVs that can be deployed, and \( z_{ik} \) is a binary association variable indicating whether \( \mathrm{UE}_i \) is served by \( \mathrm{UAV}_k \) (i.e., \( z_{ik} = 1 \) if assigned, and \( z_{ik} = 0 \) otherwise). 
The constraint in \eqref{eq:demand} ensures that the achievable data rate \( R_{ik}(\tau) \) between \( \mathrm{UE}_i \) and \( \mathrm{UAV}_k \), when associated (i.e., \( z_{ik} = 1 \)), is sufficient to satisfy the user’s traffic demand \( T_i(\tau) \). 
The constraint in \eqref{eq:capacity} guarantees that the total bandwidth allocated by \( \mathrm{UAV}_k \) to all its associated UEs does not exceed its available capacity \( B^{\text{MAX}}_k \), thereby ensuring proper bandwidth sharing among the associated users. 
The constraint in \eqref{eq:coverage} ensures that each \( \mathrm{UE}_i \) is associated with only one of the deployed UAVs, thereby guaranteeing full coverage with no user left unserved. 
The constraint in \eqref{eq:Link} ensures that \( \mathrm{UAV}_k \) can serve \( \mathrm{UE}_i \) only if it is deployed (i.e., \( a_k = 1 \)), thereby preventing associations with inactive UAVs. 
The constraints in \eqref{eq:binary} and \eqref{eq:activation} refer to the binary nature of the decision variables  $z_{ik}$ and $a_k$: \eqref{eq:binary} governs the UE-to-UAV association, while \eqref{eq:activation} determines the activation or deployment status of each UAV.

The problem formulated in \textbf{P} is a combinatorial and nonlinear problem, making it inherently NP-hard \cite{10726695}. The binary variables $z_{ik}$ and $a_k$ introduce integer programming complexity, as they define discrete association and activation decisions. Moreover, the problem involves a nonlinear constraint, i.e., the constraint in \eqref{eq:demand}, making it a mixed-integer nonlinear programming problem. Even simplified variants of this problem—such as the set covering problem (finding the smallest number of UAVs needed to cover all users) and the capacitated facility location problem (deciding where to place facilities with capacity limits to serve demands)—are well-known NP-hard problems  \cite{vazirani2001approximation}. This highlights the complexity of the problem. Consequently, solving this problem optimally becomes computationally intractable for large-scale scenarios. Therefore, it necessitates the use of heuristic or effective algorithms to efficiently obtain near-optimal solutions—an approach we detail in the following section.   


\section{Efficient Multi-UAV Traffic-Aware Deployment (EMTAD) Algorithm \label{Traffic- and Obstacle-aware Positioning Algorithm}}

To solve the problem in \textbf{P}, we propose the Efficient Multi-UAV Traffic-Aware Deployment (EMTAD) solution. EMTAD is designed to jointly minimize the number of UAVs and determine their optimal 3D positions while ensuring that all UE traffic demands are met under realistic communication constraints. The core idea of EMTAD is to leverage spatial traffic-awareness by estimating the maximum allowable distance between each UAV and its associated UE, such that the required data rate is maintained. Based on this distance threshold, we define a spherical coverage region for each UE. These regions reflect the feasible communication area around each UE, given its traffic demand and the probabilistic LoS/NLoS propagation environment.

\begin{algorithm}[ht]
\DontPrintSemicolon
\caption{Efficient Multi-UAV Traffic-Aware Deployment (EMTAD) Algorithm}
\label{alg:integrated_uav_positioning}
\footnotesize
\KwInput{
    \text{Traffic demand}~$T_i(\tau)$, in bits/s, for each \( UE_i \),
    \text{3D Position of $UE_i$}~$P_i = (x_i, y_i, z_i)$, for \( i = 1, \ldots, N \),
    \text{Maximum bandwidth per UAV}~$B^{\text{MAX}}$, in Hz,
    \text{Feasible zone for positioning UAVs}
}

\textbf{Initialize} $K_\text{max} = N$ \;
\For{$i = 1$ \KwTo $N$}{
    \text{Calculate $d_{ik}^{\text{MAX}}$ for each $UE_i$ using Equation~\ref{dmax}}\\ 
    Define sphere $S_i$ centered at $P_i$ with radius $d_{ik}^{\text{MAX}}$\;
}
\text{Compute initial feasible intersections combination $S_p$}\;

\While{$K_\text{max} \geq 1$}{
   \For{each placement of $K_\text{max}$ UAVs in $S_p$}{
        Assign $UAV_k$ (for $k = 1, \ldots, K_\text{max}$) to $S_p$\;
        \For{$i = 1$ \KwTo $N$}{
            \For{$k = 1$ \KwTo $K_\text{max}$}{
                $a_k = 1$\;
                \If{$d_{ik} \leq d_{ik}^{\text{MAX}}$}{
                    $z_{ik} = 1$\;
                }
                \Else{
                    $z_{ik} = 0$\;
                }
            }
            Check $\sum_{k=1}^{K_\text{max}} z_{ik} = 1$\;
            
            \If{all UEs covered}{
                Initialize $P_k$ at center of each $s_p$ for each $UAV_k$\;
                \For{$k = 1$ \KwTo $K_\text{max}$}{
                    \While{PSO not converged}{
                        Compute $n^{\text{MAX}}_k$ using Equation~(\ref{nmax})\;
                        Update $P_k$ using PSO\; 
                        Compute $d_{ik}(\tau)$ using Equation~(\ref{diks})\;
                        Compute $\varepsilon_{ik}^{\text{LoS}}(\tau)$ using Equation~(\ref{eq:los_prob})\;
                        Compute $\Omega_{ik}(\tau)$ using Equation~(\ref{eq:channel_gain})\;
                        Compute $R_{ik}(\tau)$ using Equation~(\ref{eq:datarate})\;
                        Check constraints (\ref{eq:demand}) and (\ref{eq:capacity})\;
                        \If{all constraints satisfied}{
                            Store solution $(a_k, z_{ik}, P_k)$\;
                        }
                        \Else{ 
                            Exit\;
                        }
                    }
                }
            }
        }
    }
    $K_\text{max} = K_\text{max} - 1$ \tcp*{Try fewer UAVs}
}
\text{Select solution with minimum $a_k$ and maximum $R_{ik}$ from stored solutions}\; 
\KwOutput{
    \begin{tabular}{ll}
    {Optimal number of UAVs} & $\sum_{k=1}^{K_\text{max}} a_k^*$ \\
    {Optimal positions} & $P_k^*$ \\
    {Association variable} & $z_{ik}^*$
    \end{tabular}
}
\end{algorithm}

Therefore, to fulfill the constraint in \eqref{eq:demand} while minimizing the objective in \eqref{eq:obj}, we define \( d_{ik}^{\text{MAX}} \), which represents the maximum distance between \( \mathrm{UE}_i \) and \( \mathrm{UAV}_k \) when \( z_{ik} = 1 \). From \eqref{eq:channel_gain}, (\ref{eq:datarate}), and \eqref{eq:demand}, \( d_{ik}^{\text{MAX}} \) is derived as follows: 

\begin{equation}\label{dmax}
d_{ik}^{\text{MAX}} = \sqrt{\frac{A P_T G_t G_r}{N_0 B_i(\tau) \left( 2^{\frac{T_i(\tau)}{B_i(\tau)}} - 1 \right)}}, 
\end{equation}
where 
$ A = K_0^{-1} \left[ \varepsilon_{ik}^{\text{LoS}}(\tau) \mu^{\text{LoS}} + \varepsilon_{ik}^{\text{NLoS}}(\tau) \mu^{\text{NLoS}} \right]^{-1} $, and $R_{ik}(\tau) = T_i(\tau)$.  
Accordingly, EMTAD creates a coverage sphere \( S_i \) of radius \( d_{ik}^{\text{MAX}} \) around each \( \mathrm{UE}_i \) to ensure that the UE's traffic demand is satisfied, where
\begin{equation}
\footnotesize
S_i = \left\{ (x,y,z) \in \mathbb{R}^3 \mid (x - x_i)^2 + (y - y_i)^2 + (z - z_i)^2 \leq (d_{ik}^{\text{MAX}})^2 \right\}
\end{equation}

The EMTAD algorithm proceeds in two stages to determine the best UAV deployment while satisfying the traffic demands and minimizing the number of UAVs, as follows. \\  

\begin{figure}
    \centering
    \includegraphics[scale=0.21]{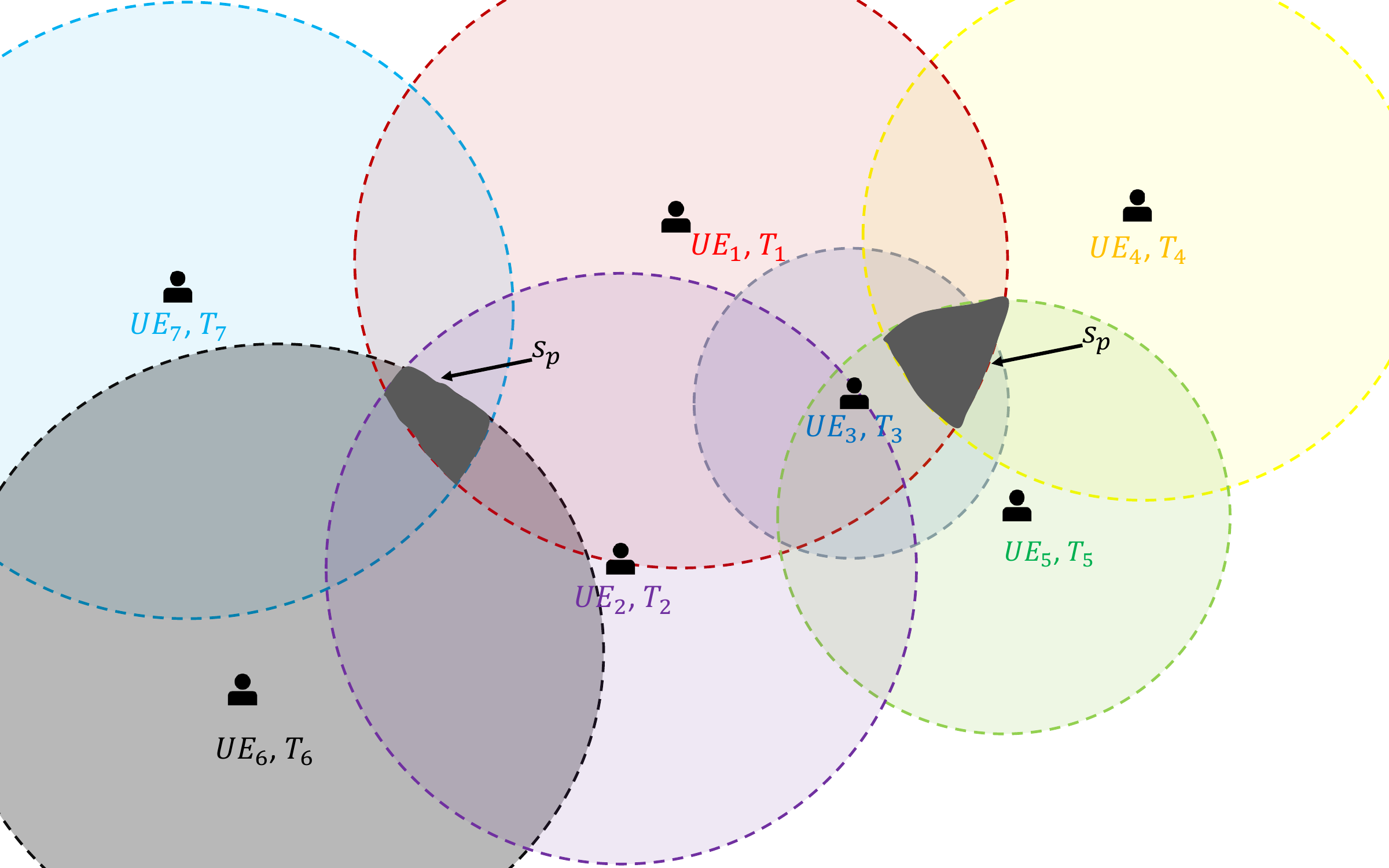}
    \caption{2D representation of the UAV positioning, which is determined by the intersection of traffic demand spheres centered at each UE. }
    \label{fig1}
\end{figure}

\textbf{i) Spatial Clustering via Coverage Intersection:} The algorithm first identifies overlapping coverage regions among UEs to determine candidate areas where a single UAV could potentially serve multiple UEs while meeting their demands. This enables the system to group UEs based on proximity and traffic compatibility, reducing unnecessary UAV deployment. To achieve this, spatial intersections between two or more of these spheres are computed. These intersections define candidate subregions \( s_p \subseteq \mathbb{R}^3 \) where a UAV can simultaneously serve multiple UEs while satisfying their respective traffic constraints. The set of all intersections is defined as:
\begin{equation}
S_p = \left\{ s_p \mid \exists I \subseteq \{1, \dots, N\}, |I| \geq 2, s_p = \bigcap_{i \in I} S_i, s_p \neq \emptyset \right\}.
\end{equation}
Then, the EMTAD algorithm identifies all intersection zones by testing UE group combinations, starting with the full set and descending to pairwise combinations. Each valid intersection zone \( s_p \) contains at least one point in 3D space that lies within all contributing spheres. Among the generated zones, the goal is to select the smallest set of intersections \( \{s_p\}_1^{a_k} \) that collectively cover all \( N \) UEs. This guarantees that the number of UAVs is minimized while fulfilling traffic constraints. The optimal set is determined by evaluating combinations of zones and identifying the smallest subset that includes all UEs. Figure \ref{fig1} shows a simple scenario that includes seven UEs. The best minimum combination requires at least two intersections \( s_p \) to cover all UEs and meet the traffic demands, where each intersection will be associated with one UAV. \\  

\textbf{ii) 3D Positioning Optimization via Particle Swarm Optimization (PSO):} 
We adopt a PSO-based metaheuristic to optimize the UAV deployment by determining an optimal position \( P_k \) within each \( s_p \). The PSO algorithm searches for the minimum set of UAVs and their corresponding 3D positions using a fitness function that jointly satisfies the following objectives:
\begin{itemize}
\item Maximize the number of UEs served per UAV, while satisfying the constraint in  (\ref{eq:capacity}), such that 
\begin{equation}\label{nmax}
n^{\text{MAX}}_k = \max \left\{ n \in \mathbb{N} \,\middle|\, \sum_{i=1}^{n} B_i(\tau) \leq B^{\text{MAX}}_k \right\}
\end{equation}
where $n^{MAX}_k$ is the maximum number of UEs that can be served by UAV $k$.   
\item Satisfy the required data rates and capacity constraints (\( d_{ik} \leq d_{ik}^{\text{MAX}} \));
\item Prioritize UAV placements that offer high LoS probability based on elevation angles and urban topology.
\end{itemize}
Once the minimum set of intersection zones is established, each intersection \( s_p \) is assigned a UAV (\(\triangleright\)~lines~9-10, Algorithm \ref{alg:integrated_uav_positioning}). 
This guarantees that the UAVs are placed in positions ensuring strong connectivity and efficient resource utilization while satisfying the QoS demands (\(\triangleright\)~lines~11-33, Algorithm \ref{alg:integrated_uav_positioning}).  
Thus, the proposed EMTAD solution balances coverage efficiency with practical deployment considerations, including spectrum sharing, user demand heterogeneity, and environmental propagation dynamics. The main steps of our solution are detailed in Algorithm~\ref{alg:integrated_uav_positioning}.

\begin{table}[ht]
    \caption{\textbf{Summary of PSO algorithm parameters.}}
    \label{tab1}
    \setlength{\tabcolsep}{3pt}
    \resizebox{0.48\textwidth}{!}{
    \begin{tabular}{p{120pt} p{140pt}}
        \hline
        \multicolumn{2}{c}{PSO Algorithm Parameters} \\  
        \hline    
        Number of particles & 30 $^{\mathrm{a}}$\\
        Maximum iterations & 100 $^{\mathrm{b}}$\\
        Position precision & 1.0 m $^{\mathrm{c}}$\\
        Inertia weight (w) & 0.7 $^{\mathrm{d}}$ \\
        LoS probability constant (urban) (c1) & 9.6 \\
        LoS probability constant (urban) (c2) & 0.28 \\
        Initial position space & Valid positions from each $s_p$ \\
        Fitness function & Aggregate Throughput\\
        LoS Threshold ($\varepsilon^{\text{Threshold}}$) & 0.9 \\
        Early stopping & After 10 iterations if a perfect solution  \\
        Position selection & Closest valid position from $s_p$ \\
      
        \hline
        \multicolumn{2}{p{240pt}}{$^{\mathrm{a}}$ Standard PSO population size balancing exploration and computational cost}\\
        \multicolumn{2}{p{240pt}}{$^{\mathrm{b}}$ Sufficient iterations for convergence while maintaining efficiency}\\
        \multicolumn{2}{p{240pt}}{$^{\mathrm{c}}$ Practical precision for UAV positioning in real-world scenarios}\\
        \multicolumn{2}{p{240pt}}{$^{\mathrm{d}}$ Balances exploration and exploitation in particle movement}\\
        \end{tabular}
      }
\end{table}

 \begin{table}[ht]
    \caption{\textbf{ NS-3 environment configuration parameters}}
    \label{tab2}
    \setlength{\tabcolsep}{3pt}
    \resizebox{0.48\textwidth}{!}{
    \begin{tabular}{p{140pt} p{120pt}}
        \hline
        \multicolumn{2}{c}{ns-3 simulation parameters} \\  
        \hline
        $f$ & 5250 MHz\\
        Guard Interval $(GI)$ & 800 $ns$\\
        Wi-Fi channel & 50 \\
        Wi-Fi Standard & IEEE 802.11ac  \\
        Channel Bandwidth & 20 MHz  \\
        Antenna Gain & 0 dBi  \\
        Tx power& 20 $dBm$\\
        Noise floor power & -85 $dBm$\\
        Remote Station Manager mechanism & IdealWifiManager  \\
        Application Traffic & UDP constant bitrate \\
        UDP Data Rate $(B_i)$ & variable based on $MCS_i$ \\
        Packet Size & 1400 bytes  \\
        \hline
        
    \end{tabular}
    }
\end{table}

\section{performance Evaluation~\label{sec:performance_evaluation}}

This section assesses the performance of the proposed EMTAD  algorithm.  
Herein, we compare EMTAD against two state-of-the-art baselines: a fixed-altitude UAV approach \cite{Alzenad20173DPlacement} and a fixed number of UEs per UAV approach \cite{Wu2018JointTrajectory, Mozaffari2016UAVD2D}, using Python scripts and ns-3 simulator. 
To assess the efficacy of the EMTAD, we conducted a comprehensive evaluation across three distinct scenarios. \textit{Scenario A} with constant venue size and number of UEs and different traffic demands, \textit{Scenario B} with constant number of UEs and traffic demands and different venue sizes, and \textit{Scenrio C} with constant venue size and traffic demands and different numbers of UEs where the UEs are randomly distributed in the venue. We consider a UAV equipped with an IEEE 802.11ac transceiver operating in the 5 GHz ISM band with a total bandwidth of $B^{MAX}=160 MHz$, enabled by channel bonding (8~$\times$~20~MHz). 802.11ac-compliant systems support this configuration and can be feasible in scenarios with relatively low interference and sufficient spectrum availability, such as rural or lightly populated areas~\cite{Mozaffari2016UAVD2D}.

\begin{figure}[t!]
    \centering
    \subfloat[Number of UAV]{
        \includegraphics[width=0.94\linewidth, height=5cm]{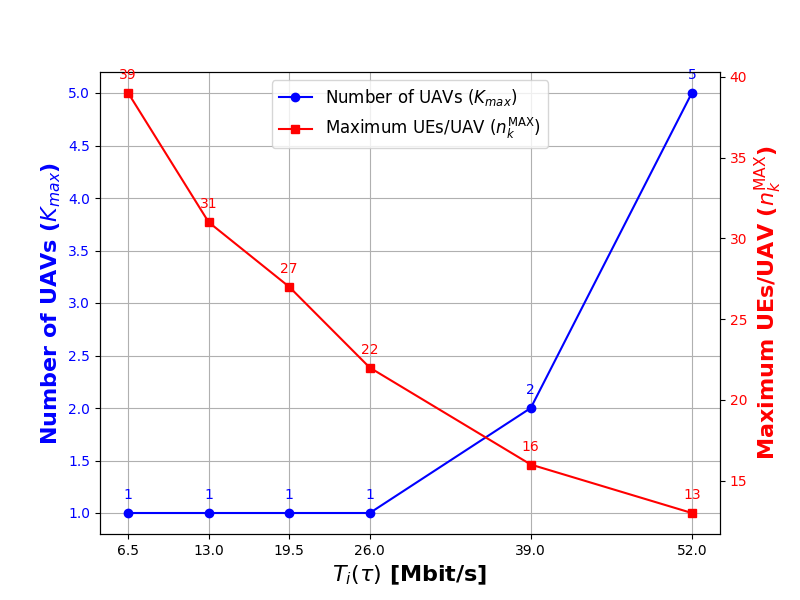}
        \label{fig2a}
    }
    \vfil
    \subfloat[Average Aggregate Throughput (Mbit/s)]{
        \includegraphics[width=0.94\linewidth, height=5cm]{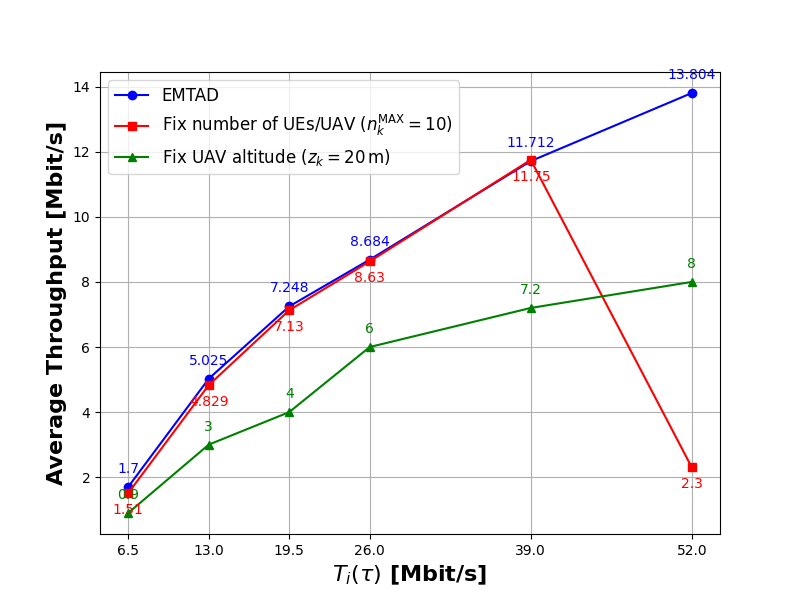}
        \label{fig2b}
    }
    \caption{Scenario A: (a) Number of UAVs and maximum UEs per UAV as a function of traffic rate \( T_i \), and (b) average aggregate throughput achieved from EMTAD compared to with fixed UAV altitude (\( z_k = 20 \, \text{m} \)) and fixed number of UEs per UAV ($n^{\text{MAX}}_k$=10).}
    \label{fig2}
\end{figure}

To evaluate the performance of EMTAD, each scenario is simulated in ns-3. Results are obtained from 30 simulation runs for each scenario. All simulations are conducted under consistent networking conditions, utilizing \texttt{SetRandomSeed()} and \texttt{RngRun = \{1, 2, \ldots, 30\}}. A summary of the parameters of the EMTAD algorithm and the simulation settings is provided in Tables~\ref{tab1} and~\ref{tab2}. A detailed description of each scenario is outlined below.

\begin{figure}[t!]
    \centering
    \subfloat[Number of UAV]{
        \includegraphics[width=0.93\linewidth, height=5cm]{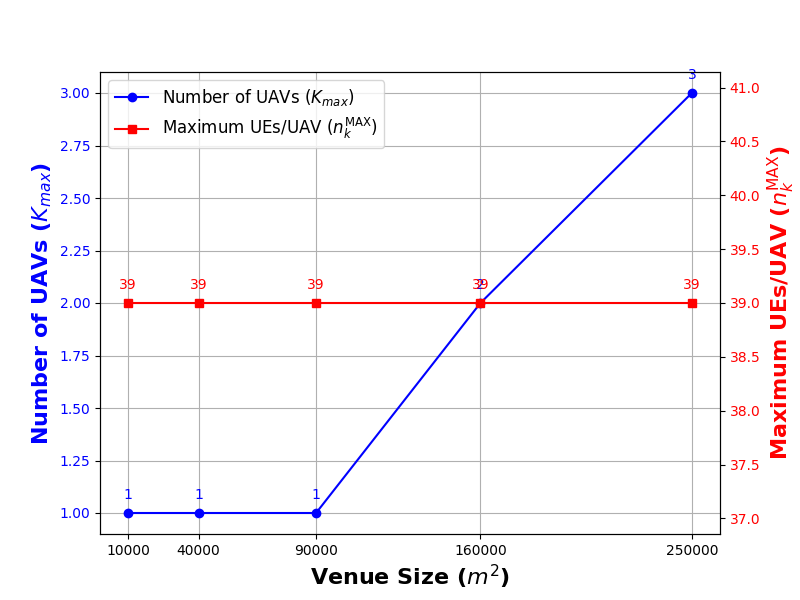}
        \label{fig3a}
    }
    \vfil
    \subfloat[Average Aggregate Throughput (Mbit/s)]{
        \includegraphics[width=0.93\linewidth, height=5cm]{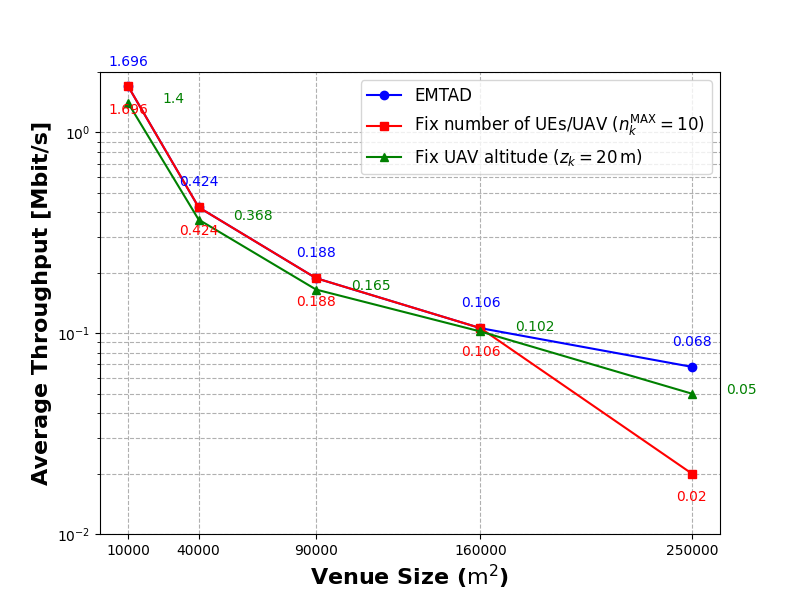}
        \label{fig3b}
    }
    \caption{Scenario B: (a) Number of UAVs and maximum UEs per UAV as a function of venue size, and (b) average aggregate throughput achieved from EMTAD compared to with fixed UAV altitude (\( z_k = 20 \, \text{m} \)) and fixed number of UEs per UAV ($n^{\text{MAX}}_k$=10).}
    \label{fig3}
\end{figure}

In \textit{Scenario A}, we consider a \(100\,\text{m} \times 100\,\text{m}\) area including 20 UEs distributed randomly, each demanding uniform traffic rates according to the Modulation and Coding Scheme (MCS) indices \cite{MCSTable91:online}. The traffic rates are set as follows: i) 6.5 Mbit/s, ii) 13 Mbit/s, iii) 19.5 Mbit/s, iv) 26 Mbit/s, v) 39 Mbit/s, and vi) 52 Mbit/s, corresponding to MCS indices 0 through 5. A guard interval of 800\,ns is applied. Figure~\ref{fig2a} illustrates the minimum number of UAVs required for each use case, considering both the accommodation of user traffic demands and the limitation of maximum UEs covered per UAV. figure \ref{fig2b} compares the average aggregate throughput received by UAVs from MTOAP with state-of-the-art, where i) a fixed number of UEs ($N_{UE}$ = 10) are associated with each UAV, and ii) a fixed altitude of UAVs ($z_k$ = 20). The results show that EMTAD outperforms both the fixed-altitude and fixed-number-of-UEs-per-UAV approaches, particularly when EMTAD proposes more UAVs. This improvement is attributed to the inability of a lower number of UAVs to accommodate the traffic demands of all UEs adequately.

In \textit{Scenario B},  20 UEs are randomly distributed in a venue of different area as follows: i) \(100\,\text{m} \times 100\,\text{m}\), ii) \(200\,\text{m} \times 200\,\text{m}\), iii) \(300\,\text{m} \times 300\,\text{m}\), iv) \(400\,\text{m} \times 400\,\text{m}\), and v) \(500\,\text{m} \times 500\,\text{m}\). Each UE demands uniform traffic rates of 6.5 Mbit/s corresponding to MCS index 0 with a guard interval of 800\,ns. The results are depicted in Figure \ref{fig3a} and \ref{fig3b}. In this scenario, EMTAD demonstrates superior performance compared to state-of-the-art approaches. For example, in areas of $100~\text{m}^2$, $200~\text{m}^2$, and $300~\text{m}^2$, EMTAD achieves the same throughput using a single UAV, whereas the benchmark approach with $N_{\text{UE}} = 10$ requires two UAVs to deliver comparable results. Furthermore, in the $500~\text{m}^2$ scenario, EMTAD proposes the deployment of three UAVs to meet the traffic demands. In contrast, the two UAVs suggested by the state-of-the-art approach are insufficient to satisfy these requirements.

\begin{figure}[t!]
    \centering
    \subfloat[Number of UAV]{
        \includegraphics[width=0.93\linewidth, height=5cm]{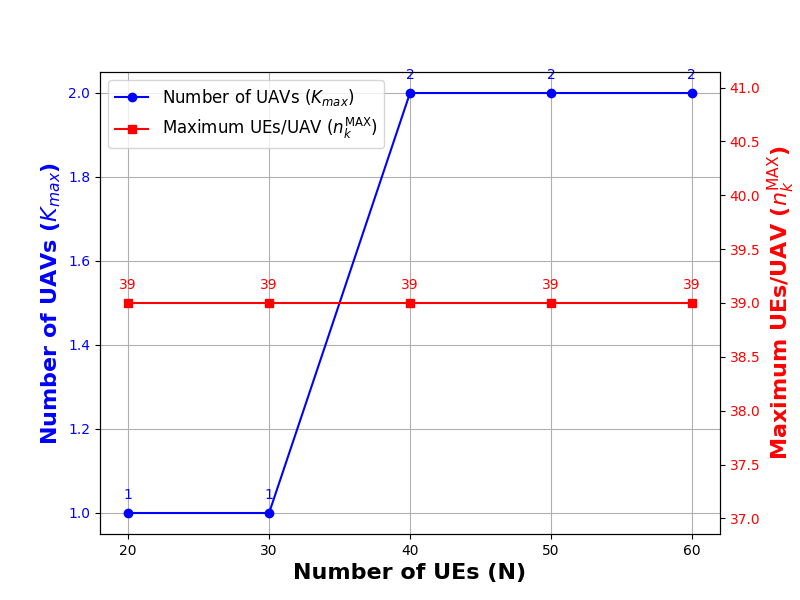}
        \label{fig4a}
    }
    \vfil
    \subfloat[Average Aggregate Throughput (Mbit/s)]{
        \includegraphics[width=0.93\linewidth, height=5cm]{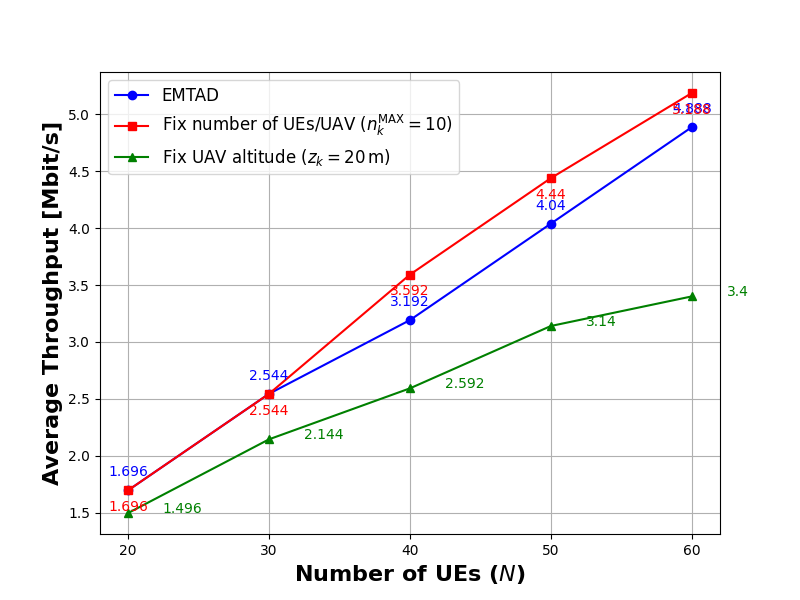}
        \label{fig4b}
    }
    \caption{Scenario C: (a) Number of UAVs and maximum UEs per UAV as a function of the number of UEs $N$, and (b) average aggregate throughput achieved from EMTAD compared to with fixed UAV altitude (\( z_k = 20 \, \text{m} \)) and fixed number of UEs per UAV ($n^{\text{MAX}}_k$=10).}
    \label{fig4}
\end{figure}

In \textit{Scenario C}, we consider a \(100\,\text{m} \times 100\,\text{m}\) area that includes different numbers of UEs distributed randomly as follows: i) 20 UEs, ii) 30 UEs, iii) 40 UEs, iv) 50 UEs, and v) 60 UEs. Each UE demands uniform traffic rates of 6.5 Mbit/s corresponding to MCS index 0 with a guard interval of 800\,ns (Figure \ref{fig4}). The results of this scenario highlight the strong performance of EMTAD, which successfully covers all UEs and accommodates their traffic demands using at most two UAVs. This is in contrast to the state-of-the-art approach, which requires up to six UAVs for the same scenario, despite offering no significant improvement in throughput.

\section{Conclusions~\label{sec:conclusions}}

This paper presented the EMTAD algorithm—a scalable and adaptive solution for optimizing UAV deployment in urban wireless networks. EMTAD addresses the critical challenge of minimizing the number of UAVs required to maintain reliable connectivity, while simultaneously ensuring that each user’s traffic demand is met.  
Through a joint optimization framework and a traffic-aware deployment strategy, EMTAD balances network load and resource usage, making it well-suited for dense and dynamic urban environments. 
By incorporating real-time traffic distribution, dynamic UE-UAV associations, and practical deployment constraints, EMTAD significantly improves spectral efficiency and LoS coverage compared to conventional static approaches.  
Future work will extend this framework to support UAV mobility, energy-aware operation, and integration with 6G-native edge intelligence for real-time decision-making in more complex scenarios.  

\balance

\bibliographystyle{IEEEtran}
\bibliography{IEEEabrv,References}

\end{document}